\RequirePackage{amsmath}
\documentclass{article}
\usepackage{cite}
\usepackage{amsmath,amssymb,amsfonts,mathtools}
\usepackage{algorithmic}
\usepackage{graphicx}
\usepackage{textcomp}
\usepackage{hyperref}
\usepackage{soul}
\usepackage{color}
\usepackage[dvipsnames]{xcolor}
\usepackage{booktabs}
\usepackage[left=1in,right=1in,top=1in,bottom=1in]{geometry}
\usepackage{pdfpages}
\usepackage[nodisplayskipstretch]{setspace}
\newcommand{\footremember}[2]{%
	\footnote{#2}
	\newcounter{#1}
	\setcounter{#1}{\value{footnote}}%
}
\newcommand{\footrecall}[1]{%
	\footnotemark[\value{#1}]%
} 

\begin{document}

\title{DiffeoRaptor: Diffeomorphic Inter-modal Image Registration using RaPTOR}

\author{Nima Masoumi \footremember{alley}{Department of Electrical and Computer Engineering, Concordia University, Montreal, Canada} \\ n\_masoum@encs.concordia.ca\and Hassan Rivaz \footrecall{alley} \\ hrivaz@ece.concordia.ca\and M. Omair Ahmad \footrecall{alley} \\ omair@ece.concordia.ca\and Yiming Xiao\footremember{trailer}{Department of Computer Science and Software Engineering, Concordia University, Montreal, Quebec, Canada}\\ yiming.xiao@concordia.ca}
\date{}


\maketitle
\begin{abstract}
~ ~ ~ ~\\
\textbf{Purpose}: Diffeomorphic image registration is essential in many medical imaging applications. Several registration algorithms of such type have been proposed, but primarily for intra-contrast alignment. Currently, efficient inter-modal/contrast diffeomorphic registration, which is vital in numerous applications, remains a challenging task.

\noindent
\textbf{Methods}: We proposed a novel inter-modal/contrast registration algorithm that leverages Robust PaTch-based cOrrelation Ratio (RaPTOR) metric to allow inter-modal/contrast image alignment and bandlimited geodesic shooting demonstrated in Fourier-Approximated Lie Algebras (FLASH) algorithm for fast diffeomorphic registration.

\noindent
\textbf{Results}: 
The proposed algorithm, named DiffeoRaptor, was validated with three public databases for the tasks of brain and abdominal image registration while comparing the results against three state-of-the-art techniques, including FLASH, NiftyReg, and Symmetric image normalization (SyN).

\noindent
\textbf{Conclusions}: Our results demonstrated that DiffeoRaptor offered comparable or better registration performance in terms of registration accuracy. Moreover, DiffeoRaptor produces smoother deformations than SyN in inter-modal and contrast registration. The code for DiffeoRaptor is publicly available at \url{https://github.com/nimamasoumi/DiffeoRaptor}.
\end{abstract}

\section{Introduction}
\label{sec:intro}
Diffeomorphic image registration allows the computation of  a smooth and invertible deformation field, and thus ensures that salient image features are not lost after image resampling with the obtained deformation fields. A key step in many clinical applications, diffeomorphic image registration can be employed in quantifying inter-subject variability of brain~\cite{b1}, studying Alzheimer's disease~\cite{alzDiff}, statistical shape analysis~\cite{revShape}, brain atlas construction~\cite{genAtlas}, and estimation of tissue deformation for surgery~\cite{b5}. 

Several studies have proposed diffeomorphic algorithms to perform intra-modal/contrast image registration. Beg \textit{et al.}~\cite{b6} implemented the Large-Deformation Diffeomorphic Metric Mapping (LDDMM) to register brain MRIs of Alzheimer's and Schizophrenia patients, but their computational cost was high. Later, many algorithms were proposed to make the computation more efficient. Vialard \textit{et al.}~\cite{b7} shortened the computational time by employing geodesic shooting to register 3D MRI scans of fetus brains. Zhang \textit{et al.}~\cite{b8} proposed Fourier-Approximated Lie Algebras (FLASH) to perform inter-subject registration of 3D brain MRIs. Similar to~\cite{b7}, they also employed geodesic shooting and improved the efficiency by performing the calculations in a band-limited space. Wu \textit{et al.}~\cite{b9} implemented cross-correlation (CC)-based LDDMM for fast brain image registration via GPU acceleration.

In general, performing diffeomorphic image registration with iterative optimization can be computationally expensive and time-consuming. Therefore, a number of deep learning (DL)-based algorithms were designed to tackle this problem~\cite{dlRev1}.
In~\cite{l2rP}, the comparison with multiple registration tasks suggests that compared with DL-based techniques, classic registration methods still have good performance and can offer satisfactory speed with the option of parallel computing.

In the last decade, several groups have attempted to design inter-modal diffeomorphic image registration techniques in various applications. Mitra \textit{et al.}~\cite{b10} proposed an inter-modal diffeomorphic algorithm to register 2D transrectal ultrasound images to MR slices. Kutten \textit{et al.}~\cite{b11} implemented the mutual information (MI)-based LDDMM on a Hamiltonian framework to register CLARITY images. Reaungamornrat \textit{et al.}~\cite{b12} proposed a MIND Demons which is based on SyN~\cite{b13}, diffeomorphic Demons~\cite{b14}, and MIND features~\cite{b15} to perform deformable MRI-CT registration for image-guided surgery. However, inter-modal image registration remains a challenging task in medical image registration. In general, the algorithms should show a certain degree of robustness against intensity inhomogeneities, noise, and image artifacts. Moreover, the algorithms should be time-efficient for real clinical applications. To address some of these requirements, Rivaz \textit{et al.}~\cite{b16} proposed RaPTOR to register 3D inter-modal images of the BITE database~\cite{b17}. Later in~\cite{b18}, an affine version of RaPTOR was used to successfully register inter-modal images of RESECT~\cite{b19} and BITE~\cite{b17} databases. Recently in~\cite{b20}, a rigid version of RaPTOR was employed to register preoperative CT and intraoperative US images of lumbar vertebrae. 

This study intends to design a diffeomorphic algorithm to perform intra- and inter-modal image registration. In~\cite{b16,b18,b20}, it was shown that RaPTOR could successfully align images with different modalities. In~\cite{b8}, it was shown that FLASH could perform computationally efficient diffeomorphic registrations compared to vector momentum LDDMM~\cite{vecMom}. However, RaPTOR and FLASH have the following drawbacks. First, RaPTOR uses B-spline as the transformation model which does not guarantee a smooth inverse transformation. Second, FLASH uses sum-of-squared differences (SSD) that is unable to directly measure the similarity between images of different modalities and contrasts~\cite{ssdIssue}. Therefore, FLASH cannot be used to perform inter-modal/contrast image registration. Third, FLASH does not use multiresolution image pyramids to tackle larger deformations which is a standard approach in many inter-modal image registration methods. Herein, we proposed DiffeoRaptor, a novel algorithm to bring together the benefits of RaPTOR and FLASH while mitigating their drawbacks. We decided to build on this similarity metric by making it diffeomorphic. Other excellent choices are normalized Gaussian fields (NGF) and MIND. FLASH framework was selected in favor of other diffeomorphic approaches. Because it is based on the well-established LDDMM framework. The performance of DiffeoRaptor was demonstrated in three applications, including 1) healthy individual MRI-to-template registration; 2) registration between Alzheimer's disease (AD) and healthy brains, as well as brain scans at different stages of AD; 3) nonlinear registration of MR and CT abdominal data. The contributions of this work are three-fold:
\begin{enumerate}
	\item Proposing a diffeomorphic image registration framework using RaPTOR.
	\item Devising inter-modal/contrast image registration with geodesic shooting in the bandlimited space of velocity fields.
	\item Employing gradient descent (GD) with momentum to improve the convergence in contrast to classical GD optimization in FLASH and RaPTOR .
\end{enumerate}

Our results show that DiffeoRaptor could achieve 1)  better alignment of brain and abdominal images compared to Mattes MI+SyN, NiftyReg~\cite{niftyreg}, and FLASH as assessed by Dice scores; 2)  smoother deformation fields compared to Mattes MI+SyN and NiftyReg in the alignment of brain MR images, and 3) comparable computation time with FLASH while performing more challenging tasks.

\section{Methodology}
\label{sec:method}
In this section, backgrounds of bandlimited space of velocity fields, bandlimited geodesic shooting, and formulation of RaPTOR metric are presented. Then, the formulation of DiffeoRaptor objective function is derived. Lastly, the optimization technique to minimize the objective function is detailed.

\subsection{Space of Bandlimited Velocity Fields}
In pairwise diffeomorphic image registration, the reference image $X\in\Omega$ and the source image $Y\in\Omega$ are given. Ideally, the objective is to find a mapping $\phi\in \rm Diff(\Omega)$ such that $X\circ \phi\approx Y$ and $Y\circ \phi^{-1}=X$.
Diffeomorphisms $\phi:\Omega\rightarrow\Omega$ are a smooth mapping that has an smooth inverse $\phi^{-1}$. The tangent vector space at the identity $id\in \rm Diff(\Omega)$ over the space of diffeomorphisms is defined as $V=T_{\rm id}\rm Diff(\Omega)$. Given $V$, the space of bandlimited velocity fields $\widetilde{V}$ was constructed and proper Lie algebra in this space was defined in~\cite{b8}. Time series $t\in \left[0, 1\right]$ of dffeomorphisms $\phi_t\in\rm Diff(\Omega)$ is created in the process of solving an ordinary differential equation (ODE). The time series of bandlimited velocity fields $\tilde{v}_t\in\widetilde{V}$ are related to $\phi^{-1}_t$ by Eq~\eqref{eq:veldiff}. 

\begin{equation}
\frac{d\phi^{-1}_t}{dt}=-D\phi^{-1}_t\cdot\iota\left(\tilde{v}_t\right)
\label{eq:veldiff}
\end{equation}

\noindent where $D$ is the derivative operator and $\iota:\widetilde{V}\rightarrow V$ is the inverse Fourier transform from the bandlimited space to the space of dense velocity fields~\cite{b8}. The geodesic shooting is the process of integrating the geodesic path of diffeomorphisms forward in time which is uniquely determined with the velocity $\tilde{v}_0$ in $t=0$. The geodesic evolution equation in the discrete Fourier space is defined in Eq~\eqref{eq:geo}.
\begin{equation}
\frac{\partial\tilde{v}_t}{\partial t}=-\widetilde{K}\bigg[(\widetilde{D}\tilde{v})^T\star\tilde{m}_t+\widetilde{\Gamma}(\tilde{m}_t\otimes\tilde{v}_t)\bigg]
\label{eq:geo}
\end{equation}
\noindent where $K$ is the smoothing operator which is the inverse of the differential operator $L$. There is an in-depth discussion of possible choices of $L$ in~\cite{b6,millerGeo,joshiDiff}. In this paper, it is set $L=(-\alpha\Delta+I)^c$ similar to~\cite{b6,b8} where $\Delta$ is the Laplacian operator. $\widetilde{K}$ is the smoothing operator in the bandlimited space~\cite{b8}, $\star$ is the truncated auto-correlation, $\widetilde{\Gamma}$ is the discrete divergence, $\tilde{m_t}=\widetilde{L}\tilde{v}_t$ is the momentum, $\widetilde{L}$ is the representation of $L$ in the frequency domain, $\otimes$ denotes the tensor product, and $\widetilde{D}$ is an operator that computes the spatial gradient in the bandlimited Fourier space~\cite{b8}. 

\subsection{Geodesic Shooting in the Bandlimited Space}
\label{subsec:flash}
By setting the geodesic shooting as the constraint of the cost function, it does not require calculating the velocity fields $\tilde{v}_t$ and diffeomorphisms $\phi_t$ in a dense time grid and it suffices to calculate the initial velocity $\tilde{v}_0\in\widetilde{V}$. The cost function for FLASH was defined as Eq.~\eqref{eq:flash}.

\begin{equation}
E(\tilde{v}_0)=\frac{1}{2\sigma^2}\big\|Y\circ\phi^{-1}_1-X\big\|^2+\langle\widetilde{L}\tilde{v}_0,\tilde{v}_0\rangle,\quad\rm s.t. \,Eq.~\eqref{eq:geo}
\label{eq:flash}
\end{equation}

\noindent where $\sigma$ is the noise variance, $\|\cdot\|$ is the norm operator in the space $\Omega$, $\widetilde{L}$ is the inverse of $\widetilde{K}$, and $\langle,\rangle$ is the inner-product in the space $\widetilde{V}$~\cite{b8}. Gradient of the energy function $E$ can be calculated as in Eq.~\eqref{eq:flashgrad} for the minimization of cost.

\begin{equation}
\nabla_{\tilde{v}_1}E=\nu\bigg(-K\bigg(\frac{1}{\sigma^2}(Y\circ\phi^{-1}_t-X)\cdot\nabla(Y\circ\phi^{-1}_1)\bigg)\bigg)
\label{eq:flashgrad}
\end{equation}\noindent 
where $\nu:V\rightarrow\widetilde{V}$ is the projection mapping to the bandlimited space of velocity fields and $K$ is the smoothing operator. 

\subsection{RaPTOR}
\label{subsec:raptor}
%
One possible choice for the similarity metric is the Correlation Ratio (CR)~\cite{rocheCR}. For challenging inter-modal image registration tasks, calculation of CR needs to be robust and possibly time-efficient. RaPTOR is a dissimilarity metric that is based on CR~\cite{b16} and addresses the shortcomings of CR~\cite{rocheCR}. RaPTOR and its derivative can be calculated as in Eq.~\eqref{eq:rappatchR}. It calculates CR in local patches $\Theta$. Instead of calculating the iso-sets of $X$, the histogram of $X$ over $N_b$ bins are calculated and then Parzen windowing was applied to make the bins continuous and differentiable.

\begin{subequations}
\begin{align}
1-\eta(Y|X)&=\frac{1}{N\sigma^2}\bigg(\sum_{i=1}^{N}y^{2}_i - \sum_{j=1}^{N_b}N_j\mu_j^{2}\bigg) \\
\mu_j&=\frac{\sum_{i=1}^{N}\lambda_{ij}y_i}{N_j}, N_j=\sum_{i}\lambda_{ij} \\
\mathrm{RaPTOR}(Y,X)&=\Psi(Y,X)=\frac{1}{N_p}\sum_{i=1}^{N_p}(1-\eta(Y|X;\Theta_i)) \\
\nabla_{\varphi}\Psi&=\frac{\partial \Psi}{\partial \varphi}=\frac{\partial \phi}{\partial \varphi}\cdot\frac{\partial Y}{\partial \phi}\cdot\frac{\partial \Psi}{\partial Y} \\
\begin{split}
\frac{\partial (1-\eta)}{\partial y_i}=\frac{2}{N\sigma^2}\bigg(y_i-\lambda_{i,j-1}\mu_{j-1}-\lambda_{ij}\mu_j-\\
\frac{1}{(N-1)\sigma^2}(y_i-\mu)\bigg(\sum_{a=1}^{N}y_a^{2}-\sum_{c=1}^{N_b}N_c\mu_c^{2}\bigg)\bigg)
\label{eq:rapdercrsimpleR}
\end{split}
\end{align}
\label{eq:rappatchR}
\end{subequations}

\noindent where $N$ is the number of pixels in a image patch $\Theta_i$, $\sigma^2=\mathrm{Var}[Y;\Theta_i]$ is the variance of a patch $i$ in $Y$, $y_i$ is the intensity of sample $i$ in image $Y$, let $j$ and $j-1$ be the closest bins to sample $x_i$ (intensity of sample $i$ in $X$) then according to its distance to these bins centers, $\lambda_{ij}$ is the linear contribution of $x_i$ to the bin $j$, $N_p$ is the number of patches, $\varphi$ is the parameter of transformation $\phi$, and $\mu=E[Y]$ is the average value of $Y$. $\eta(Y|X)$ can measure the functional dependence between the input images. When there is no functional dependence $\eta(Y|X)=0$ and when  $\eta(Y|X)=1$ there is a deterministic relationship between X and Y. Calculating gradient of RaPTOR analytically enables efficient minimization of the dissimilarity metric using gradient-based optimization and employing the outlier suppression technique elaborated in~\cite{b16}.

\subsection{DiffeoRaptor}
\label{subsec:diffrap}
The energy function in Eq.~\eqref{eq:flash} can be generalized to the form in Eq.~\eqref{eq:general}.

\begin{equation}
E(\tilde{v}_0)=\mathrm{dist}\big(Y\circ\phi^{-1}_1,X\big)+\langle\widetilde{L}\tilde{v}_0,\tilde{v}_0\rangle,\quad\rm s.t. \,Eq.~\eqref{eq:geo}
\label{eq:general}
\end{equation}

\noindent where $dist(,)$ is a normalized distance function or a dissimilarity function. DiffeoRaptor is the cost function in the form of Eq.~\eqref{eq:general} with the RaPTOR defined in Eq.~\eqref{eq:rappatchR} as the dissimilarity function. So it takes the form in Eq.~\ref{eq:diffrap}.

\begin{equation}
E(\tilde{v}_0)=\Psi\big(Y\circ\phi^{-1}_1,X\big)+\langle\widetilde{L}\tilde{v}_0,\tilde{v}_0\rangle,\quad\rm s.t. \,Eq.~\eqref{eq:geo}
\label{eq:diffrap}
\end{equation}

Eq.~\eqref{eq:flashgrad} is no longer valid for Eq.~\eqref{eq:diffrap} and the gradient of cost function needs to be calculated for the optimization. A similar approach to~\cite{b6} is taken to calculate $\partial_u\Psi$, the variation of cost in Eq.~\eqref{eq:diffrap} with respect to the velocity $u=D\phi^{-1}_1$ which is obtained by taking the derivative of $\phi^{-1}_1$.

Given the fact that we are working with image intensities in a grid according to Eq.~\eqref{eq:rappatchR}, the variation of energy $\partial_uE$ takes the form $\partial_uE=\langle\nabla_u E,u\rangle_{V_g}$ and therefore $\partial_u\Psi=\langle\nabla_u \Psi,u\rangle_{V_g}$. The inner-product $\langle,\rangle_{V_g}$ calculation is over a finite grid ($V_g$ is the space of velocities where the inner-product $\langle,\rangle_{V_g}$ is taken). To calculate the Gateaux derivative of cost in Eq.~\eqref{eq:diffrap}, one is required to derive $\partial_u\Psi$ first as in Eq.~\eqref{eq:rapH1}.


\begin{equation}
\partial_u\Psi=\langle\frac{\partial \Psi}{\partial Y}\cdot\nabla(Y\circ\phi^{-1}_1),u\rangle_{V_g}
\label{eq:rapH1}
\end{equation}

%
Detailed derivation of  Eq.~\eqref{eq:rapH1} is presented in Section S5 of Supplementary Material. Eq.~\eqref{eq:rapH1} indicates that $\nabla_{u}\Psi=\frac{\partial \Psi}{\partial Y}\cdot\nabla(Y\circ\phi^{-1}_1)$ which is known and can be calculated using the Eq.~\eqref{eq:rapdercrsimpleR}. By similar calculations to~\cite{b6} and~\cite{b8}, the gradient of cost can be written as Eq.~\eqref{eq:diffgrad}.
\begin{equation}
\nabla_{\tilde{v}_1}E=\nu\bigg(-K\bigg(\frac{\partial \Psi}{\partial Y}\cdot\nabla(Y\circ\phi^{-1}_1)\bigg)\bigg)
\label{eq:diffgrad}
\end{equation}
\noindent
the gradient in Eq.~\eqref{eq:diffgrad} for the velocity in $t=1$ can be used to find the gradient $\nabla_{\tilde{v}_0}E$ in $t=0$ with the reduced adjoint Jacobi field in bandlimited velocity fields elaborated in~\cite{b8}. This process is called backward integration. To minimize the cost in Eq.~\ref{eq:diffrap}, forward integration of Eq.~\ref{eq:geo} is used to find the velocity in $t=1$. Then $\nabla_{\tilde{v}_0}E$ is used in GD with momentum optimization to update the velocity. Finally, Eq.~\ref{eq:veldiff} is used to calculate diffeomorphisms. Since similar process was used in~\cite{b8} to calculate diffeomorphisms, the diffeomorphic registration is guaranteed. The employment of multi-resolution pyramid, gradient descent with momentum, and implementation details of DiffeoRaptor can be found in the Supplementary Materials.

\section{Experiments and Results}
\label{sec:results}
DiffeoRaptor was validated on three public datasets: IXI (\url{http://brain-development.org/ixi-dataset}), OASIS3~\cite{oasis3}, and The Cancer Imaging Archive (TCIA) MR-CT abdominal data~\cite{tcia}. It is compared against Mattes MI+SyN, which is available in Advanced Normalization Tools (ANTs)~\cite{ants} and NiftyReg~\cite{niftyreg} (using the normalized mutual information (NMI) as the similarity metric), as well as in several tasks with FLASH. Dice scores of overlapping regions are used as evaluation metrics. The default parameters for NiftyReg with the GD optimization produced the best results for us. Mattes MI+SyN is a diffeomorphic algorithm which uses Mattes MI as the similarity metric and models the deformation fields with SyN, and is suitable for inter-modal/contrast image registration. The parameters for Mattes MI+SyN were tuned such that it produced the optimal results. The number of bins for MI was set to $32$ and the gradient step, the update field variance, and total field variance were set to 0.5, 3, and 0.5 for SyN respectively.

\subsection{Pre-processing of Brain MRI}
\label{subsec:prepro}

\begin{table}
	\centering
	\caption{Abbreviation of subcortical structures which were automatically labelled in the segmentation of brain volumes using volBrain~\cite{volbrain}.}
	\label{table:abbrv}
	\setlength{\tabcolsep}{3pt}
	\resizebox{0.3\columnwidth}{!}{
		\begin{tabular}{lc}
			\toprule
			Subcortical Structure& Abbreviation \\
			\midrule
			Left/Right Ventricle & LV/RV \\	
			Left/Right Caudate & LC/RC \\
			Left/Right Putamen & LP/RP \\
			Left/Right Thalamus & LT/RP \\
			Left/Right Globus Pallidus & LGP/RGP \\
			Left/Right Hippocampus & LH/RH \\
			Left/Right Amygdala & LA/RA \\
			Left/Right Accumbens & LAC/RAC \\
			\bottomrule
	\end{tabular}}
\end{table}

\begin{figure}[!h]
	\centering
	\includegraphics[width=0.5\columnwidth]{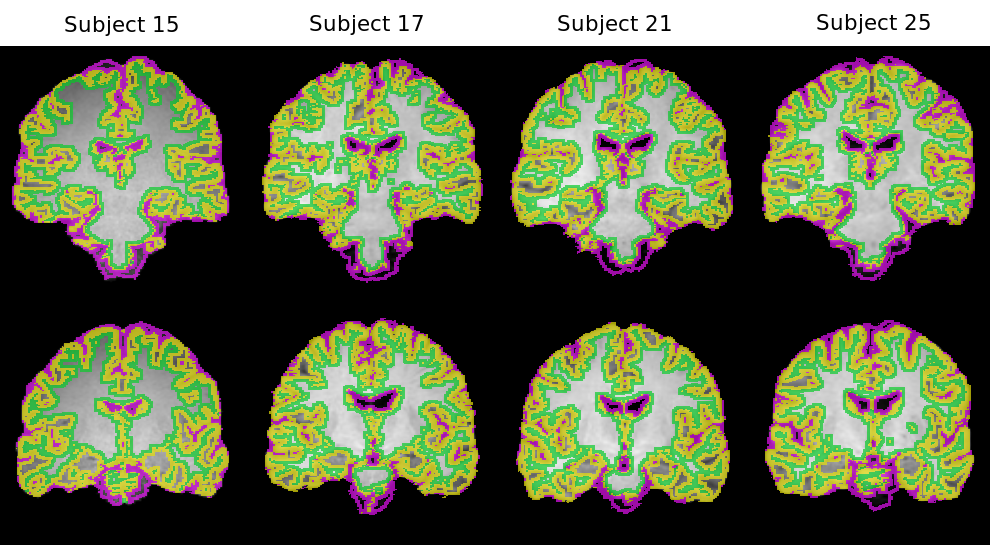}
	\caption{Coronal view of two slices (rows) of four different IXI dataset subjects (columns). The images are overlaid by the segmentation of CSF, GM, and WM. The large variability of structures across subjects require a deformable registration.}
	\label{fig:ixi}
\end{figure}

Brain MR images of the IXI and OASIS3 datasets, were first skull-stripped using nonlocal intracranial cavity extraction~\cite{beast}. For each case, the extracted brain was carefully inspected. Then, two types of segmentations were generated for each volume using the volBrain algorithm~\cite{volbrain} so that Dice scores can be used to evaluate registration accuracy. Here, in the first one, brain tissues are classified into Cerebrospinal Fluid (CSF), Gray Matter (GM), and White Matter (WM). The second type of segmentation consists of 16 subcortical structures which are abbreviated in Table~\ref{table:abbrv}.  Lastly, the volumes were affinely registered using ANTs with Mattes MI as the metric (see Fig.~\ref{fig:ixi}).

\subsection{IXI Dataset: Inter-subject Registration}
\label{subsec:ixi}

\begin{table}
	\centering
	\caption{Dice score (mean$\pm$sd) evaluation of T1-T1, T1-T2, and T1-PD registrations of IXI dataset for DiffeoRaptor, Mattes MI+SyN, FLASH, and NiftyReg in overlapping regions of brain tissues and sixteen subcortical structures.}
	\label{table:ixi}
	\setlength{\tabcolsep}{2pt}
	\resizebox{0.7\columnwidth}{!}{
		\begin{tabular}{lccccc}
			\toprule
			Task and Evaluation Region & Affine only & DiffeoRaptor & Mattes MI+SyN & FLASH & NiftyReg\\
			\midrule
			T1-T1 Brain Tissues & $0.62\pm 0.03$ & $0.72\pm 0.04$ & $0.72\pm 0.04$ & $0.64\pm 0.04$ & $0.67\pm 0.03$\\
			T1-T1 Subcortical Structures & $0.67\pm 0.06$ & $0.78\pm 0.03$ & $0.78\pm 0.04$ & $0.67\pm 0.06$ & $0.74\pm 0.05$\\		
			T1-T2 Brain Tissues & $0.62\pm 0.03$ & $0.67\pm 0.04$ & $0.67\pm 0.04$ & $0.64\pm 0.03$ & $0.64\pm 0.04$\\
			T1-T2 Subcortical Structures  & $0.67\pm 0.06$& $0.71\pm 0.06$ & $0.71\pm 0.06$ & $0.66\pm 0.05$ & $0.65\pm 0.05$\\		
			T1-PD Brain Tissues & $0.62\pm 0.03$ & $0.67\pm 0.04$ & $0.67\pm 0.04$ & $0.64\pm 0.03$ & $0.64\pm 0.04$\\
			T1-PD Subcortical Structures & $0.67\pm 0.06$& $0.71\pm 0.05$ & $0.69\pm 0.06$ & $0.67\pm 0.06$ & $0.68\pm 0.06$\\ \bottomrule	
	\end{tabular}}
\end{table}

Twenty young adult subjects ($age<30yo$) of the IXI dataset were selected randomly. Given the fact that the IXI datatset offers T1w, T2w, and PDw for each subject, three different tasks were designed, including T1-T1, T1-T2, and T1-PD registrations. T1w MRI scans of three subjects (Subject 15, 17, and 21) were randomly selected as the reference volume and the rest are set as the source volumes for inter-subject registration (in total $3\times 19=57$ registrations). The results of Dice score evaluation are summarized in Table~\ref{table:ixi}, which shows that DiffeoRaptor, Mattes MI+SyN, and NiftyReg could successfully align volumes in each task whereas FLASH underperformed in terms of Dice scores in intra-contrast tasks and failed in inter-contrast tasks as expected. It can also be seen that DiffeoRaptor in general did better than Mattes MI+SyN. 

\subsection{IXI Dataset: Subject-to-template Registration}
\label{subsec:icbm}

\begin{table}
	\centering
	\caption{Dice score (mean$\pm$sd) evaluation of ICBM152-T1, ICBM152-T2, and ICBM152-PD registrations of IXI dataset for DiffeoRaptor, Mattes MI+SyN, FLASH, and NiftyReg in overlapping regions of brain tissues and sixteen subcortical structures.}
	\label{table:icbm}
	\setlength{\tabcolsep}{3pt}
	\resizebox{0.7\columnwidth}{!}{
		\begin{tabular}{lccccc}
			\toprule
			Task and Evaluation Regions & Affine only & DiffeoRaptor & Mattes MI+SyN & FLASH & NiftyReg\\
			\midrule
			ICBM152-T1 Brain Tissues & $0.62\pm 0.04$ & $0.68\pm 0.04$ & $0.70\pm 0.04$ & $0.63\pm 0.04$ & $0.71\pm 0.05$\\
			ICBM152-T1 Subcortical Structures & $0.70\pm 0.06$ & $0.80\pm 0.02$ & $0.78\pm 0.04$ & $0.71\pm 0.04$ & $0.74\pm 0.07$ \\		
			ICBM152-T2 Brain Tissues & $0.62\pm 0.04$ & $0.65\pm 0.05$ & $0.67\pm 0.06$ & $0.62\pm 0.04$ & $0.65\pm 0.05$\\
			ICBM152-T2 Subcortical Structures & $0.70\pm 0.06$& $0.76\pm 0.04$ & $0.76\pm 0.08$ & $0.67\pm 0.06)$ & $0.73\pm 0.08$\\		
			ICBM152-PD Brain Tissues & $0.62\pm 0.04$ & $0.66\pm 0.04$ & $0.66\pm 0.05$ & $0.63\pm 0.04$ & $0.64\pm 0.05$\\
			ICBM152-PD Subcortical Structures & $0.70\pm 0.06$& $0.75\pm 0.05$ & $0.73\pm 0.07$ & $0.70\pm 0.05$ & $0.73\pm 0.07$ \\	
			\bottomrule	
	\end{tabular}}
\end{table}

Given the IXI subjects in Section~\ref{subsec:ixi}, the volumes are set as the source volumes and they were registered to the T1w ICBM152 template~\cite{icbm}. Here, the template is set as the reference volume and similar tasks were performed as Section~\ref{subsec:ixi} for subject-to-template registration. The results are summarized in Table~\ref{table:icbm}, which shows that DiffeoRaptor, Mattes MI+SyN, and NiftyReg could successfully align volumes in each task while DiffeoRaptor in general did better than Mattes MI+SyN and NiftyReg, especially in alignment of subcortical structures.

\begin{figure}[!h]
	\centering
	\includegraphics[width=0.8\columnwidth]{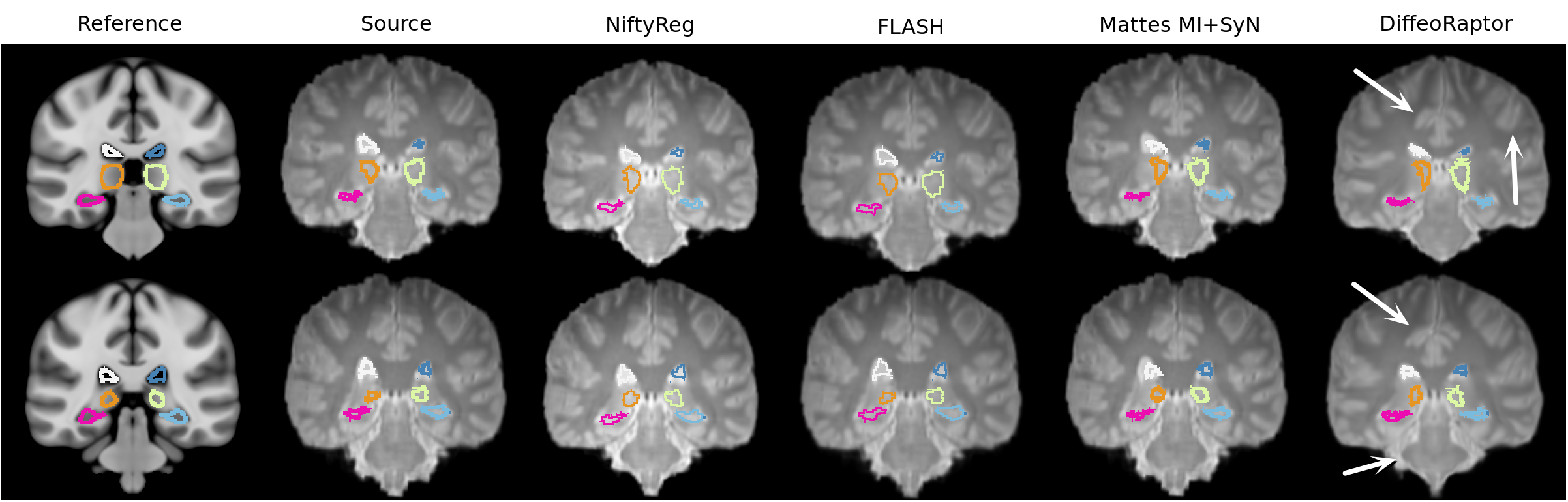}
	\caption{From the left to right: coronal slices of the ICBM152 (reference volume), the PDw source volume of the IXI dataset, result of NiftyReg, FLASH, Mattes MI+SyN, and DiffeoRaptor respectively. Rows show different coronal views. Subcortical structural segmentations are shown in colored contours. 	Arrows are pointing to the regions where the image alignments are more visible.}
	\label{fig:icbm}
\end{figure}

Figure ~\ref{fig:icbm} demonstrates two coronal views of registration results. The subcortical structures are shown in the figure as colored outlines. DiffeoRaptor show better alignment of slices and anatomical structures compared to other methods. The cerebrum shape with DiffeoRaptor registration looks closer to the ICBM152 template than other methods.

\subsection{OASIS3 Dataset: Intra- and Inter-subject Registration}
\label{subsec:oasis3}
The OASIS3 dataset consists of subjects intended for investigating Alzheimer's disease (AD)~\cite{oasis3}. Twenty AD patients from this dataset were randomly selected with matching T1w and T2w MRIs. In the first sub-task, intra-contrast intra-subject registration was performed for brain scans obtained at different stages of AD progression, where the T1w volume at the baseline was set as the reference and the T1w image from the latest session ($>$ 6 months apart) with visible atrophy was registered to the reference. This sub-task represents the need in neuroimage analysis for tracking disease-related anatomical changes. The results are included in Table S1 of the Supplementary materials.

\begin{table}
	\centering
	\caption{Dice score evaluation (mean$\pm$sd) of T1-T2 inter-subject registration of IXI data with OASIS3 data for DiffeoRaptor, Mattes MI+SyN, and NiftyReg in overlapping regions of brain tissues and sixteen subcortical structures.}
	\label{table:oasis3}
	\setlength{\tabcolsep}{3pt}
	\resizebox{0.7\columnwidth}{!}{
		\begin{tabular}{lcccc}
			\toprule
			Task and Evaluation Regions & Affine only & DiffeoRaptor & Mattes MI+SyN & NiftyReg\\
			\midrule				
			T1-T2 Brain Tissues Mean & $0.51\pm 0.13$ & $0.61\pm 0.07$ & $0.54\pm 0.16$ & $0.56\pm 0.16$\\
			T1-T2 Subcortical Structures & $0.56\pm 0.17$& $0.71\pm 0.11$ & $0.63\pm 0.21$ & $0.59\pm 0.18$\\ \bottomrule			
	\end{tabular}}
\end{table}

\begin{figure}[!h]
	\centering
	\includegraphics[width=0.7\columnwidth]{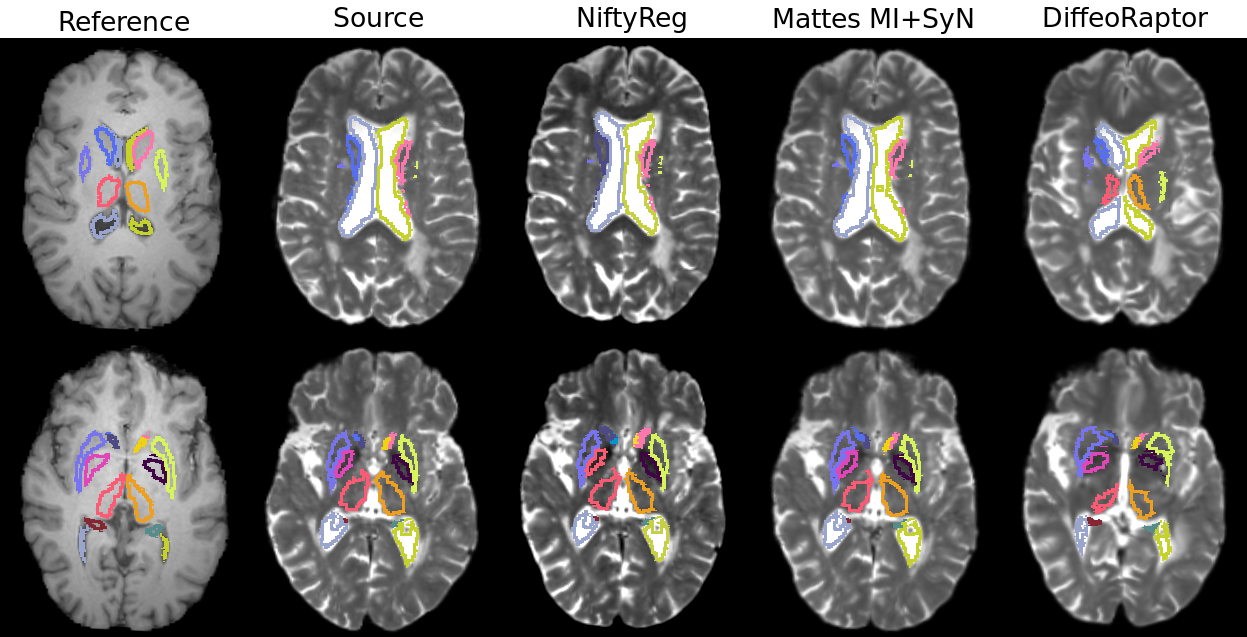}
	\caption{From the left to right: axial slices of the T1w reference volume from the IXI dataset, the T2w MRI source volume of the OASIS3 dataset, result of NiftyReg, Mattes MI+SyN, and DiffeoRaptor respectively. Rows show different axial views. Subcortical segmentations are shown in colored contours.}
	\label{fig:oasis}
\end{figure}

In the second sub-task, T1w MRIs of four young healthy adults of the IXI datatset in Section~\ref{subsec:ixi} were used as the references and the T2w MRI scans of the latest session for each subject from the OASIS3 dataset were set as the source volumes, resulting in $4\times 20=80$ registrations. This way, we defined a more challenging, inter-contrast, inter-subject, and inter-dataset task to better compare DiffeoRaptor with Mattes MI+SyN and NiftyReg.The results of T1-T2 registrations are summarized in Table 4, where DiffeoRaptor outperformed Mattes MI+SyN and NiftyReg. Note that FLASH was not included in these experiments because it continuously failed to perform inter-contrast registration. In Fig.~\ref{fig:oasis}, it can be seen that DiffeoRaptor has improved the alignment of subcortical structures and ventricles better than Mattes MI+SyN and NiftyReg.

\subsection{TCIA Abdominal MR-CT Intra-subject Registration}
\label{subsec:l2r}
The TCIA dataset contains eight subjects. Each subject has a T1w MRI scan and CT scan (with deformation) of the abdomens. The manual segmentations of the liver, spleen, left kidney, and right kidney are provided by the Learn2Reg organizers (\url{https://learn2reg.grand-challenge.org}). By setting the MRI scan for each subject as the reference volume, CT scans were aligned to perform intra-subject registrations. The deformable registration for MR-CT of these subjects are required because the images were taken in different time points, with different modalities, and misalignments due to patient movement, respiration, and etc. The results are summarized in Table~\ref{table:l2r}.

\begin{table}
	\centering
	\caption{Dice score (mean$\pm$sd) evaluation of MR-CT intra-subject registration for TCIA abdominal data using DiffeoRaptor, RaPTOR~\cite{b16}, Mattes MI+SyN, and NiftyReg.}
	\label{table:l2r}
	\setlength{\tabcolsep}{3pt}	
	\resizebox{0.5\columnwidth}{!}{	
	\begin{tabular}{lcccc}
		\toprule
		Evaluation Regions & DiffeoRaptor & RaPTOR & Mattes MI+SyN & NiftyReg\\
		\midrule
		Liver & $0.81\pm 0.07$ & $0.81\pm 0.09$ & $0.80\pm 0.10$ & $0.79\pm 0.11$ \\
		Spleen & $0.71\pm 0.13$ & $0.71\pm 0.16$ & $0.69\pm 0.10$ & $0.71\pm 0.13$ \\		
		Left Kidney& $0.70\pm 0.15$ & $0.71\pm 0.15$ & $0.68\pm 0.17$ & $0.65\pm 0.22$ \\
		Right Kidney& $0.70\pm 0.19$ & $0.69\pm 0.19$ & $0.67\pm 0.17$ & $0.65\pm 0.22$ \\				
		\midrule
		Average & $0.78\pm 0.10$ & $0.77\pm 0.10$ & $0.77\pm 0.11$ & $0.76\pm 0.13$\\
		\bottomrule
	\end{tabular}}
\end{table}

\begin{figure}[!h]
	\centering
	\includegraphics[width=0.8\columnwidth]{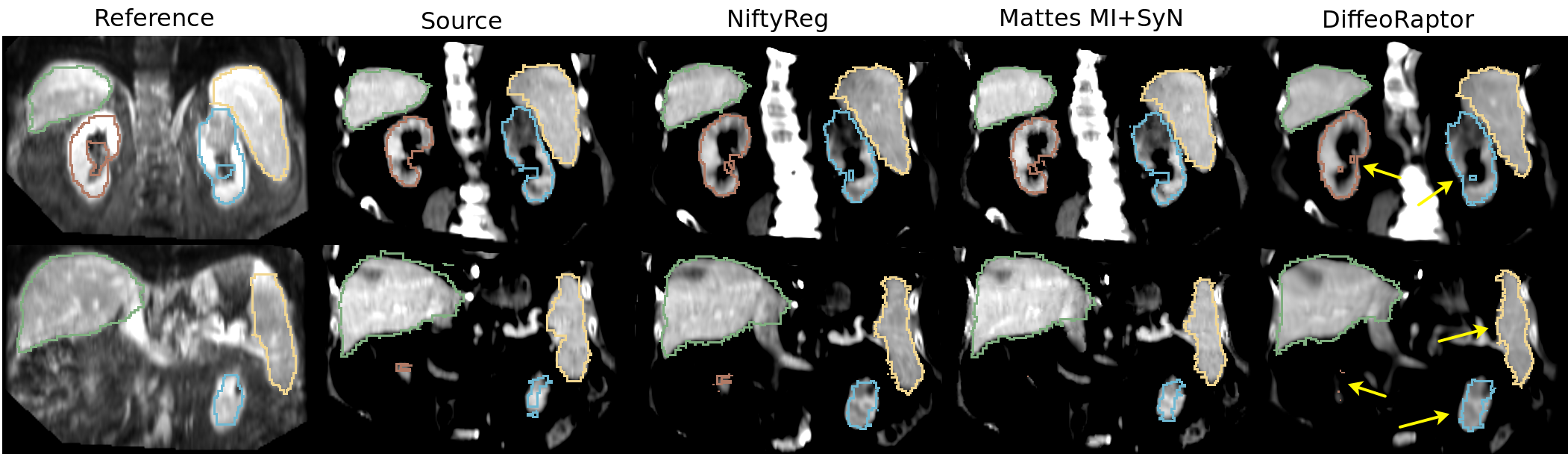}
	\caption{From left to right: coronal slices of Subject 7's MRI (reference volume), the corresponding CT source volume, results of NiftyReg, Mattes MI+SyN, DiffeoRaptor, and NiftyReg respectively. Rows show different slices of volumes. Segmentations of key organs are shown with colored contours. Arrows are pointing to the regions where the image alignments are more visible.}
	\label{fig:l2r}
\end{figure}

Given the fact that the initial affine registration achieved mean Dice score of $0.72\pm0.10$, Table~\ref{table:l2r} shows DiffeoRaptor, RaPTOR~\cite{b16}, Mattes MI+SyN, and NiftyReg could successfully improve the image alignment. Besides, DiffeoRaptor outperformed Mattes MI+SyN and NiftyReg in alignment of all the targeted regions. Note that two subjects didn't have the segmentation of the right kidney and thus they were excluded from the Mean Dice calculation of Table~\ref{table:l2r}.  In Fig. ~\ref{fig:l2r}, it can be seen that compared to the affine registration, Mattes MI+SyN and DiffeoRaptor show improvement in alignment of segmented organs. However, DiffeoRaptor shows better alignment of organs compared to Mattes MI+SyN and NiftyReg.

\subsection{Cumulative Results}
\label{subsec:cumu}
\begin{table}
	\centering
	\caption{Dice scores (mean$\pm$sd) of cumulative results for DiffeoRaptor,  Mattes MI+SyN, and NiftyReg in overlapping subcortical structures. The p-values from ANOVA are shown for each anatomical structure.}
	\label{table:cumu}
	\setlength{\tabcolsep}{6pt}
	\resizebox{0.6\columnwidth}{!}{
		\begin{tabular}{ccccc}
			\toprule
			Evaluation Regions & DiffeoRaptor & Mattes MI+SyN & NiftyReg & p-value\\
			\midrule
			LV & $0.65\pm 0.14$ & $0.60\pm 0.18$ & $0.58\pm 0.18$ & $1.21\times 10^{-5}$\\
			RV & $0.64\pm 0.13$ & $0.59\pm 0.17$ & $0.57\pm 0.16$ & $1.60\times 10^{-6}$ \\		
			LC & $0.73\pm 0.11$ & $0.70\pm 0.17$ & $0.66\pm 0.16$ & $4.30\times 10^{-6}$ \\
			RC & $0.72\pm 0.11$ & $0.68\pm 0.19$ & $0.64\pm 0.17$ & $1.15\times 10^{-5}$ \\
			LP & $0.79\pm 0.08$ & $0.75\pm 0.14$ & $0.73\pm 0.14$ & $4.90\times 10^{-7}$ \\
			RP & $0.78\pm 0.08$ & $0.73\pm 0.15$ & $0.70\pm 0.16$ & $2.33\times 10^{-11}$ \\
			LT & $0.80\pm 0.09$ & $0.78\pm 0.16$ & $0.73\pm 0.15$ & $1.58\times 10^{-7}$ \\
			RT & $0.78\pm 0.09$ & $0.77\pm 0.15$ & $0.71\pm 0.15$ & $6.56\times 10^{-8}$ \\
			LGP & $0.70\pm 0.10$ & $0.65\pm 0.14$ & $0.60\pm 0.17$ & $1.04\times 10^{-13}$ \\
			RGP & $0.68\pm 0.10$ & $0.64\pm 0.14$ & $0.57\pm 0.17$ & $2.44\times 10^{-15}$ \\
			LH & $0.69\pm 0.09$ & $0.65\pm 0.14$ & $0.64\pm 0.14$ & $7.31\times 10^{-5}$ \\
			RH & $0.73\pm 0.09$ & $0.69\pm 0.14$ & $0.67\pm 0.14$ & $8.56\times 10^{-8}$ \\
			LA & $0.60\pm 0.15$ & $0.55\pm 0.18$ & $0.53\pm 0.18$ & $3.37\times 10^{-5}$ \\
			RA & $0.60\pm 0.14$ & $0.57\pm 0.17$ & $0.53\pm 0.17$ & $7.06\times 10^{-6}$ \\
			LAC & $0.50\pm 0.17$ & $0.43\pm 0.21$ & $0.37\pm 0.22$ & $1.22\times 10^{-11}$ \\
			RAC & $0.46\pm (0.18$ & $0.45\pm 0.20$ & $0.36\pm 0.21$ & $7.70\times 10^{-9}$ \\
			\midrule
			Average & $0.72\pm 0.08$ & $0.68\pm 0.14$ & $0.65\pm 0.13$ & $1.14\times 10^{-8}$\\
			\bottomrule
	\end{tabular}}
\end{table}

\begin{figure}[!h]
	\centering
	\includegraphics[width=0.3\columnwidth]{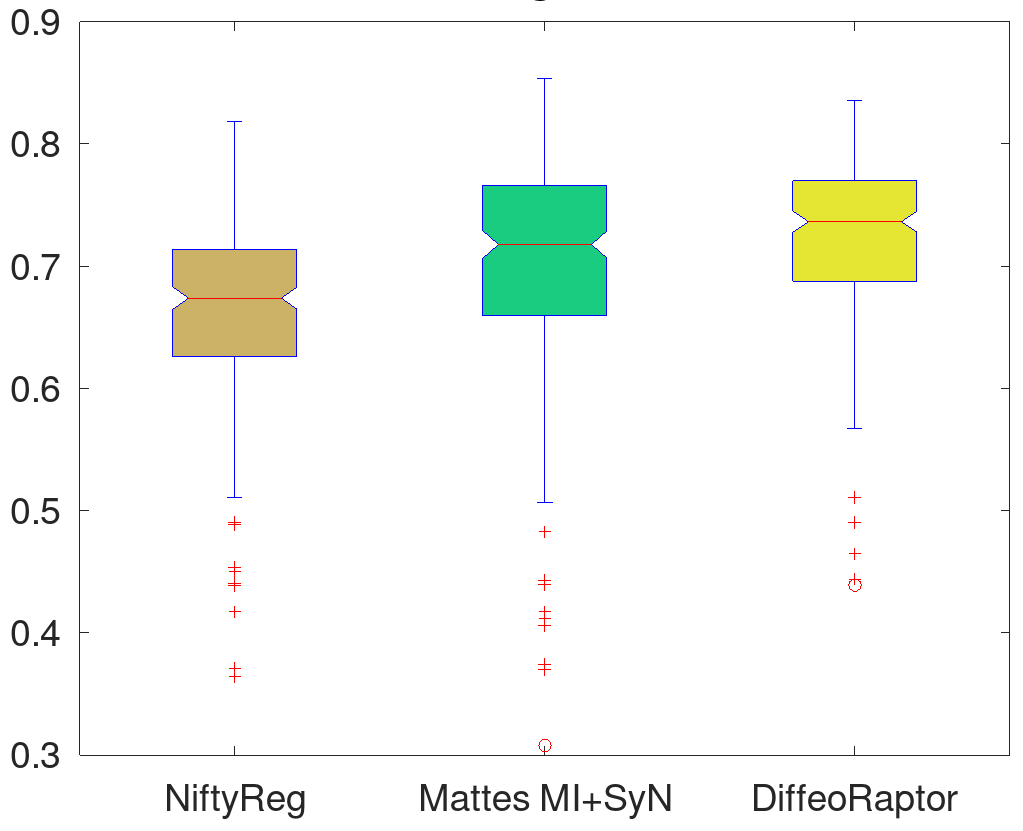}
	\caption{The box plots of average Dice score for the total of 291 brain image registrations. DiffeoRaptor has a higher mean and lower std with fewer outliers.}
	\label{fig:bp}
\end{figure}

Given the inter-contrast registration results (total 291) in Section~\ref{subsec:ixi},~\ref{subsec:icbm}, and~\ref{subsec:oasis3} for brain structures, the mean Dice scores and the associated p-values from comparing the three methods using the one-way analysis of variance (ANOVA) were listed for the sixteen subcortical structures in Table~\ref{table:cumu}. Furthermore, post-hoc multiple comparison (Tukey-Kramer) tests were performed to reveal the performance of the methods (Table~\ref{table:tucky}). With the statistical tests, we confirm that DiffeoRaptor outperforms the rest in terms of Dice scores for aligning each subcortical region, as well as the mean Dice score ($p<0.05$). It is worth mentioning that the average mean Dice is $0.63\pm0.12$ for the affine registration. To better visualize the results for the last row of Table~\ref{table:cumu}, the box plots of average Dice scores over all evaluation regions are demonstrated in Fig.~\ref{fig:bp}.

\begin{table}
	\centering
		\caption{Post-hoc multiple comparison (Tukey-Kramer) tests of DiffeoRaptor against Mattes MI+SyN and NiftyReg for the average Dice in overlapping subcortical structures. DiffeoRaptor results are better than Mattes MI+SyN and NiftyReg ($p<0.05$).}
		\label{table:tucky}
		\setlength{\tabcolsep}{6pt}
		\resizebox{0.35\columnwidth}{!}{
			\begin{tabular}{cc}
				\toprule
				Methods &  p-value\\
				\midrule
				DiffeoRaptor vs Mattes MI+SyN & $1.92\times 10^{-2}$\\
				DiffeoRaptor vs NiftyReg & $3.62\times 10^{-6}$ \\					
				\bottomrule
	\end{tabular}}
\end{table}

\subsection{Deformation Smoothness Analysis}
\label{subsec:jac}

\begin{figure}[!h]
	\centering
	\includegraphics[width=\linewidth]{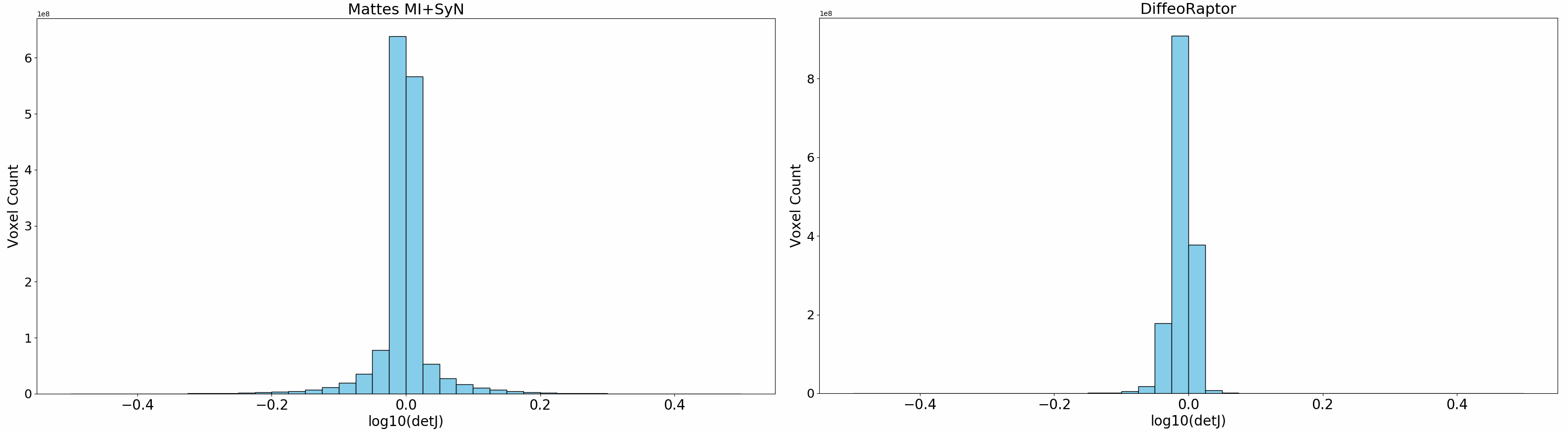}
	\caption{The logarithm of determinant of Jacobian $log_{10}(det(J))$ was calculated for each voxel of the deformation field. Then they were accumulated in bins for Mattes MI+SyN and DiffeoRaptor.}
	\label{fig:jac}
\end{figure}

With the cumulative results in Section~\ref{subsec:cumu}, in each registration, the determinant of Jacobian $J$ was calculated for each voxel of the deformation field. Figure ~\ref{fig:jac} shows $log_{10}(det(J_\phi))$ for each voxel that they were accumulated in bins for DiffeoRaptor and Mattes MI+SyN. For example, the bin centred at the origin means no deformation, bins with the negative centres show contraction, and bins with positive centre show expansion. The further the bin from the centre, the more deformation the bin represents. From the experiments, we observed that the number of non-zero samples is similar across DiffeoRaptor, Mattes MI+SyN, and NiftyReg. However, DiffeoRaptor has fewer samples far from the central bin (Fig.~\ref{fig:jac}), and generates smoother deformations than Mattes MI+SyN and NiftyReg, as shown in Table S2 of the Supplementary Materials. The determinants of Jacobians for DiffeoRaptor, Mattes MI+SyN, and NiftyReg are visualized in Fig. S2 of Supplementary Material. For the ablation study, the deformation smoothness of DiffeoRaptor, RaPTOR~\cite{b16}, Mattes MI+SyN, and NiftyReg are comapred in the TCIA abdominal dataset and the results are summarized in Table S3 of the Supplementary Materials.

\section{Discussions}
\label{sec:disc}

When RaPTOR is employed as the similarity metric, it may require additional parameter-tuning. This motivates more advanced optimization technique rather than the classical GD to minimize the cost function. This was shown and explored in~\cite{b18} and~\cite{b20}. For DiffeoRaptor, the parameter settings were mostly the default values from RaPTOR and FLASH as elaborated previsouly. However, for the cases where affine registration fails to perform good initial alignments, we should be careful in choosing the step size for the gradient update and the maximum number of iterations. The average computational times were calculated for DiffeoRaptor and FLASH on a single core of a 6 core Linux Mint system for 10 T1-T1 brain MRI registrations with the image size of $176\times 256\times 256$ voxels. The mean computational time per registration of DiffeoRaptor ($384.50\pm0.01s$) is comparable to that of FLASH ($416.14\pm0.01s$).
It should be noted that there are issues with using surrogates such as tissue overlap to evaluate the performance of registration methods~\cite{manSeg}, as outlined in more detail in the Supplementary Materials.

\section{Conclusion}
\label{sec:conc}
We present DiffeoRaptor, a diffeomorphic inter-modal/contrast image registration algorithm based on RaPTOR and geodesic shooting in bandlimited space. The algorithm is validated on several different applications. Compared with FLASH, Mattes MI+SyN, and NiftyReg, it achieves comparable or better results. In addition, DiffeoRaptor offers smoother deformation fields than Mattes MI+SyN and NiftyReg.

\section*{Acknowledgment}

This work was supported in part by the Natural Sciences and Engineering Research Council of Canada (NSERC) and in part by the Regroupment Strategique en Microelectronique du Quebec. Data were provided in part by OASIS-3: Principal Investigators: T. Benzinger, D. Marcus, J. Morris; NIH P50 AG00561, P30 NS09857781, P01 AG026276, P01 AG003991, R01 AG043434, UL1 TR000448, R01 EB009352. AV-45 doses were provided by Avid Radiopharmaceuticals, a wholly owned subsidiary of Eli Lilly.

\section{Compliance with Ethical Standards}
\textbf{Conflict of interest} The authors declare that there is no conflict of interest.

\noindent
\textbf{Ethical standard} All procedures perfomed in studies involving human participants were in accordance with the ethical standards of the institutional and/or national research committee and with the 1964 Declaration of Helsinki and its later amendments or comparable ethical standards.

\noindent
\textbf{Informed consent} Informed consent was obtained from all participants included in the study.

\begin{spacing}{0.87}

\end{spacing}


\section*{Supplementary material}

# Supplementary Materials for DiffeoRaptor: Diffeomorphic Inter-modal Image Registration using RaPTOR

Nima Masoumi [*]
n_masoum@encs.concordia.ca

Hassan Rivaz [*]
hrivaz@ece.concordia.ca

M. Omair Ahmad [*]
omair@ece.concordia.ca

Yiming Xiao[†]
yiming.xiao@concordia.ca

## S1  Mutli-resolution Pyramid and Gradient Descent with Momentum

For DiffeoRaptor, the Gradient Descent (GD) with momentum was used to break the zig-zag pattern of GD and reduce the chance of trapping in a local minimum. Thus, the convergence for DiffeoRaptor could be possibly better than Fourier-Approximated Lie Algebras (FLASH) and Robust PaTch-based cOrrelation Ratio (RaPTOR). The update equation in iteration $i$ for GD with momentum is outlined in Eq. 1. In Eq. 1, $p_i$ is the momentum parameter, and $(\tilde{v}_0)_i$ is the velocity field in iteration $i$. Setting $p_i = \frac{0.1}{1-(0.9)^{i+1}}$ produces best results for us.

$$(\tilde{v}_0)_{i+1} = (\tilde{v}_0)_i - \epsilon \nabla_{\tilde{v}_0} E + p_i((\tilde{v}_0)_i - (\tilde{v}_0)_{i-1}) \tag{1}$$

Multi-resolution image pyramid is employed often in image registration techniques to improve convergence of the algorithm. However, FLASH does not implement this technique. In DiffeoRaptor, it was implemented for up to three levels of image pyramids to produce larger deformations. Fig. S1 shows image energy term ($\Psi$), total energy (E), and Mean Square Error (MSE) between the template and source image over the iterations for three image pyramid levels. In this test case, there were maximum ten iterations needed for the convergence. Since RaPTOR and followingly DiffeoRaptor use random patches to calculate $\Psi$, the quantities over the iterations are not strictly decreasing. Moreover, for the Level 0, MSE is increasing after the 5th iteration which does not necessarily mean that the image alignment has been deteriorated.

## S2  Implementation Details

Parameter settings for DiffeoRaptor are similar to what is elaborated in RaPTOR [1] and FLASH [2]. The patch size of $7 \times 7 \times 7$ and number of patches of 3000, 2000, and 200 for the finest to the coarse level respectively, produce the best result according to the following parameter search experiment. In this experiment, six subjects of IXI dataset are selected which are not used (hold-out validation set) in the image registrations experiments of Section 3.2. $\binom{6}{2} = 15$ image registrations are performed by selecting two distinct subjects each time. Table S1 shows five different parameter setups and mean Dice score of overlapping regions. Setup 1 has better mean Dice score than the other setups. Setup 1 and 5 have similar mean Dice scores, however, the image registrations using Setup 5 takes more execution time due to the selection of more patches.

Given the discussion on the size of bandlimited space in [3], the bandlimited space size is set to $16 \times 16 \times 16$. However, the maximum number of iterations are smaller for each level of image pyramids compared to FLASH's maximum number of iterations in the full resolution. Note that the registration in the full resolution has the greatest impact on the overall computational time. Typically in our experiments, the number of iterations are fewer than 30 for each level of image pyramids.

[*]Department of Electrical and Computer Engineering, Concordia University, Montreal, Canada
[†]Department of Computer Science and Software Engineering, Concordia University, Montreal, Quebec, Canada

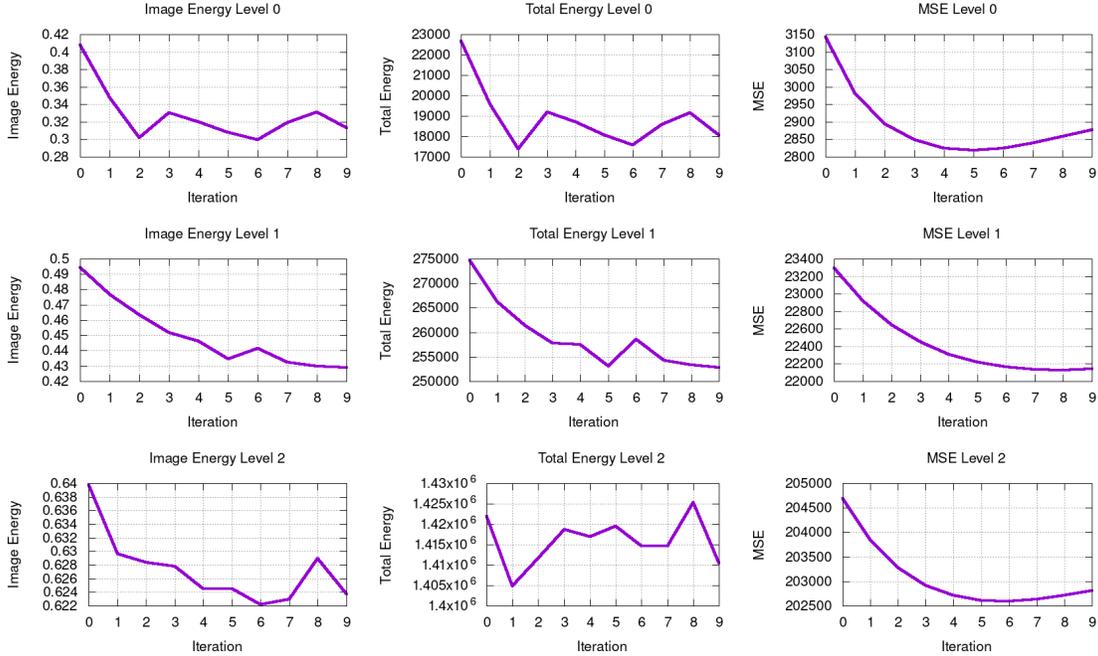

Figure S1: Image energy term ($\Psi$), tatal energy (E), and Mean Square Error (MSE) between the template and source image over the iterations for three image pyramid levels. In this test case which is a T1 image to a template registration, maximum five iterations were needed for the convergence,

Table S1: Mean Dice score evaluation of 15 T1-T1 inter-subject registrations of hold-out validation set of IXI data for DiffeoRaptor in overlapping regions of brain tissues and sixteen subcortical structures for 5 parameter setups. Note that the number of patches are for the finest to the coarse level respectively.

| Setup | Patch size | Number of patches | Dice - brain tissue | Dice - subcortical structure |
|---|---|---|---|---|
| 1 | $7 \times 7 \times 7$ | 3000, 2000, 200 | $0.70 \pm 0.01$ | $0.83 \pm 0.02$ |
| 2 | $5 \times 5 \times 5$ | 3000, 2000, 200 | $0.70 \pm 0.01$ | $0.82 \pm 0.02$ |
| 3 | $9 \times 9 \times 9$ | 3000, 2000, 200 | $0.70 \pm 0.01$ | $0.82 \pm 0.02$ |
| 4 | $7 \times 7 \times 7$ | 1500, 1000, 100 | $0.69 \pm 0.01$ | $0.82 \pm 0.02$ |
| 5 | $7 \times 7 \times 7$ | 5000, 3000, 400 | $0.70 \pm 0.01$ | $0.83 \pm 0.02$ |

Similar to Rivaz *et al.* [1], by using a test dataset, we noticed that setting $Y$ as the moving image and $X$ as the fixed image produced the best results for us. Note that in all the registrations, $X$ is the reference T1w image and $Y$ can be T1w, T2w, PDw, or CT.

## S3 T1w-T1w Intra-subject Registration of OASIS3 Dataset

The results of intra-subject registration of OASIS3 data were summarized in Table S2. In Table S2, it can be seen that for the first task (T1-T1 intra-subject registration), DiffeoRaptor, Mattes MI+SyN, and NiftyReg improved the Dice score by 0.01. The reason is that the affine registration aligned the images such that there was less room for improvement.

## S4 Further Validation of Deformation Smoothness Analysis

To further validate the smoothness of deformation fields, the quantity $\sum |\log_{10}(det(J))|$ was calculated where the summation is over all the samples available and the results are summarized in Table. S3. It can be seen that the summation values in Mattes MI+SyN and NiftyReg are higher than DiffeoRaptor. This suggests

Table S2: Mean Dice score evaluation of T1-T1 intra-subject registration of OASIS3 data for DiffeoRaptor, Mattes MI+SyN, and NiftyReg in overlapping regions of brain tissues and sixteen subcortical structures.

| Task and Evaluation Regions | Affine only | DiffeoRaptor | Mattes MI+SyN | NiftyReg |
|---|---|---|---|---|
| T1-T1 Brain Tissues | $0.83 \pm 0.02$ | $0.84 \pm 0.01$ | $0.83 \pm 0.01$ | $0.84 \pm 0.02$ |
| T1-T1 Subcortical Structures | $0.87 \pm 0.02$ | $0.88 \pm 0.01$ | $0.88 \pm 0.01$ | $0.88 \pm 0.01$ |

Table S3: Comparison of deformation smoothness of DiffeoRaptor, Mattes MI+SyN, and NiftyReg by the determinant of Jacobians of deformation fields ($J_\phi$).

| | Method | $\log_{10}(\det(J_\phi)) \neq 0$ | $\sum |\log_{10}(\det J_\phi)|$ | $\det(J_\phi) \leq 0$ |
|---|---|---|---|---|
| (i) | Mattes MI+SyN | $1.49 \times 10^8$ | $2.61 \times 10^7$ | 9038 |
| (ii) | NiftyReg | $1.49 \times 10^8$ | $3.62 \times 10^7$ | 47048 |
| (iii) | DiffeoRaptor | $1.49 \times 10^8$ | $1.95 \times 10^7$ | 0 |
| | Difference of (i) and (iii) | 9038 | $6.58 \times 10^6$ | 9038 |
| | Difference of (ii) and (iii) | 47048 | $1.67 \times 10^7$ | 47048 |

that DiffeoRaptor generates smoother deformations while having generally better performance. The other way to compare the deformation fields is counting the number of voxels in which $\det(J_\phi) \leq 0$. If $\det(J_\phi) \leq 0$, it means the mapping of that voxel was not diffeomorphic. Last column of Table. S3 shows that Mattes MI+SyN and NiftyReg had many voxels with this property, whereas DiffeoRaptor did not have any voxels in which $\det(J_\phi) \leq 0$.

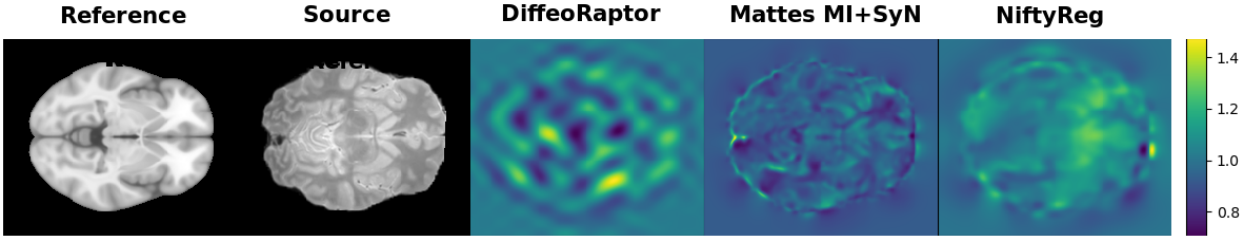

Figure S2: The visualization of Jacobian determinants for DiffeoRaptor, Mattes MI+SyN, and NiftyReg. DiffeoRaptor shows smoother deformation compared to other methods.

Figure S2 shows the visualization of Jacobian determinants for DiffeoRaptor, Mattes MI+SyN, and NiftyReg. The registration is performed between ICBM152 T1w and Subject 15 of IXI dataset (PDw). For this registration, the number of negative and zero Jacobian determinants ($\det(J_\phi) \leq 0$) are zero for all the cases. However, the value of $\sum |\log_{10}(\det J_\phi)|$ for DiffeoRaptor, Mattes MI+SyN, and NiftyReg are 12954.1, 204088.1, and 360751.5 respectively. This indicates that the deformation by DiffeoRaptor is smoother than the other methods.

Table. S4 presents the comparison of DiffeoRaptor, RaPTOR [1], Mattes MI+SyN, and NiftyReg in terms of deformation smoothness for TCIA abdominal MR-CT intra-subject registration. The quantity $\sum |\log_{10}(\det(J_\phi))|$ is calculated where the summation is over all the samples available. For the ablation study, DiffeoRaptor suggests smoother deformation compared to RaPTOR [1]. NiftyReg generated 76118 samples with $\det(J_\phi) \leq 0$ which means the deformations are not diffeomorphic. While DiffeoRaptor outperformed other methods in terms of average Dice, Mattes MI+SyN generated smoother deformations compared to other methods.

## S5 Deriving Variation of RaPTOR

The variation of RaPTOR can be calculated using the Eq. 2 by perturbing the diffeomorphism $\phi_1^{-1}$. The diffeomorphism is perturbed by the velocity $u = D\phi_1^{-1}$ which is obtained by taking the derivative of $\phi_1^{-1}$.

3

Table S4: Comparison of deformation smoothness of DiffeoRaptor, RaPTOR [1], Mattes MI+SyN, and NiftyReg by the determinant of Jacobians of deformation fields ($J_\phi$) for TCIA abdominal MR-CT intra-subject registration.

|      | Method         | $\log_{10}(\det(J_\phi)) \neq 0$ | $\sum |\log_{10}(\det J_\phi)|$ | $\det(J_\phi) \leq 0$ |
|------|----------------|----------------------------------|----------------------------------|----------------------|
| (i)  | Mattes MI+SyN  | $4.71 \times 10^7$               | $2.39 \times 10^6$               | 0                    |
| (ii) | NiftyReg       | $4.71 \times 10^7$               | $1.55 \times 10^6$               | 76118                |
| (iii)| RaPTOR         | $5.11 \times 10^7$               | $4.61 \times 10^7$               | 0                    |
| (iv) | DiffeoRaptor   | $5.11 \times 10^7$               | $3.75 \times 10^7$               | 0                    |
|      | Difference of (i) and (iv)   | $3.95 \times 10^6$ | $3.51 \times 10^7$ | 0     |
|      | Difference of (ii) and (iv)  | $4.01 \times 10^6$ | $3.59 \times 10^7$ | 76118 |
|      | Difference of (iii) and (iv) | 0                  | $8.62 \times 10^6$ | 0     |

$$\begin{aligned}
\partial_u \Psi =& \frac{1}{N_p} \sum_{k=1}^{N_p} \partial_u (1 - \eta(Y \circ \phi_1^{-1} | X; \Theta_k)) = \\
& \frac{1}{N_p} \sum_{k=1}^{N_p} \lim_{\epsilon \to 0} \frac{1}{\epsilon} \Big( \eta(Y \circ \phi_1^{-1} | X; \Theta_k) - \eta(Y \circ (\phi_1^{-1} + \epsilon u) | X; \Theta_k) \Big) = \\
& \frac{1}{N_p} \sum_{k=1}^{N_p} \lim_{\epsilon \to 0} \frac{1}{\epsilon} \Bigg( \frac{1}{N\sigma^2} \Big( \sum_{i=1}^{N} y_i^2 - \sum_{c=1}^{N_b} N_c \mu_c^2 \Big) \\
& - \frac{N-1}{N \sum_{i=1}^{N} (z_i(\epsilon) - \mu)^2} \Big( \sum_{i=1}^{N} (z_i(\epsilon))^2 - \frac{1}{N_j} \sum_{i=1}^{N} \lambda_{ij} (z_i(\epsilon))^2 \\
& - \frac{1}{N_{j-1}} \sum_{i=1}^{N} \lambda_{ij-1}(z_i(\epsilon))^2 + \frac{1}{N_j} \sum_{i=1}^{N} \lambda_{ij} y_i^2 + \frac{1}{N_{j-1}} \sum_{i=1}^{N} \lambda_{ij-1} y_i^2 \Big) \Bigg) = \\
& \Bigg( \frac{1}{N_p} \sum_{k=1}^{N_p} \frac{2}{N\sigma^2} \Big( \sum_{i=1}^{N} y_i h_i u_i - \sum_{i=1}^{N} \lambda_{i,j-1} \mu_{j-1} h_i u_i - \\
& \sum_{i=1}^{N} \lambda_{ij} \mu_j h_i u_i - \frac{1}{(N-1)\sigma^2} \Big( \sum_{i=1}^{N} (y_i - \mu) h_i \Big) \Big( \sum_{i=1}^{N} y_i^2 - \sum_{c=1}^{N_b} N_c \mu_c^2 \Big) \Big) \Bigg) \\
=& \langle \frac{\partial \Psi}{\partial Y} \cdot \nabla(Y \circ \phi_1^{-1}), u \rangle_{V_g}
\end{aligned} \qquad (2)$$

where $z(\epsilon) = Y \circ (\phi_1^{-1} + \epsilon u)$, $h = \nabla(Y \circ \phi_1^{-1})$, and $\nabla(Y \circ \phi_1^{-1})u = \lim_{\epsilon \to 0} \frac{1}{\epsilon} \Big( Y \circ (\phi_1^{-1} + \epsilon u) - Y \circ \phi_1^{-1} \Big)$ is used in the fourth equality. The term $\frac{\partial \Psi}{\partial Y}$ is a vector with $N$ elements. Eq. 3 shows element $i$th of the vector using the notation $\{\frac{\partial \Psi}{\partial Y}\}_{i=1}^{N}$.

$$\left\{ \frac{\partial \Psi}{\partial Y} \right\}_{i=1}^{N} = \frac{1}{N_p} \sum_{k=1}^{N_p} \frac{2}{N\sigma^2} \left( y_i - \lambda_{i,j-1} \mu_{j-1} - \lambda_{ij} \mu_j - \frac{1}{(N-1)\sigma^2} \Big( y_i - \mu \Big) \Big( \sum_{a=1}^{N} y_a^2 - \sum_{c=1}^{N_b} N_c \mu_c^2 \Big) \right) \qquad (3)$$

## S6 Limitations

In the work of Rohlfing *et al.* [4], it was shown that although similarity metric and automatically generated tissue overlaps are commonly used for validation of image registration accuracy, they can be unreliable. They

investigated two possible solutions for this problem: 1) utilization of localized labels and 2) using manually generated segmentations. For our experiments, the registration accuracy is validated on 16 subcortical regions. Our validation using subcortical regions showed the agreement across the results on tissue overlaps, qualitative validation, and our similarity metric (RaPTOR). Similar approaches were used for validation of image registration in most of the current registration techniques (e.g. VoxelMorph [6]). However, we admit that these validation approaches (tissue overlap and similarity metrics) have their limitations. When resources are available, validation using an established manually labelled dataset and multi-modal ground truths (e.g. addition of matching anatomical landmarks) will form better assessments.

The segmentations of 16 subcortical structures are generated by the automated algorithm (volBrain [5]). The algorithm uses non-local means patch-based segmentation and only affine registration to align the target MRI and the atlas library. The algorithm was compared against the widely used brain tissue/structure segmentation software, FreeSurfer [7] and FIRST [8] in in terms of Dice coefficients on a manually labelled dataset [5]. The average Dice scores of $0.91 \pm 0.05$, $0.73 \pm 0.11$, and $0.77 \pm 0.10$ were reported for volBrain, FreeSurfer, and FIRST, respectively [5], showing excellent performance of volBrain. Worth to mention that most of the current learning-based registration works (e.g. VoxelMorph [6]) use automatic segmentation, such as FreeSurfer for validation purposes due to limited resources for manually labeled datasets.

In Section 3.3, the T1w ICBM152 template was used in the automated algorithm volBrain to generate tissue maps and segmentation of subcortical structures. Due to partial volume effects from averaging several subjects, the generated segmentations may not be as accurate as expected. This could possibly be mitigated by manual segmentations of individual subject's brain MRI that contributed towards the T1w ICBM152, and then thresholding the averaged tissue maps. Unfortunately, this is not feasible due to limitation of resources and public availability of the data.

The Task 1 training dataset of Learn2Reg challenge is used in our MR-CT registration in Section 3.5. However, the leader-board only shows the result for the test dataset, which can be downloaded from the challenge website (https://learn2reg.grand-challenge.org/Learn2Reg2021/) but it does not include the associated image segmentations for validation purposes. Therefore, we decided to compare our results with those official participants of the Learn2reg challenge, whose verified results are published in the associated challenge summary paper [9]. In the report, the top three methods for Task 1 achieved the average Dice score of 0.78, 0.76, and 0.75 respectively. Given our results in Table 5 of the manuscript, the average Dice score of 0.78 was achieved. Therefore, DiffeoRaptor potentially has a similar performance to the top methods that participated in the challenge.

	
\end{document}